\documentclass[preprint,12pt]{elsarticle}

\usepackage{amssymb}

\newproof{proof}{Proof}

\journal{Discrete Applied Mathematics}

\usepackage{amsmath}
\usepackage{mathrsfs}
\usepackage{graphicx}
\usepackage{color}
\usepackage[all]{xy}

\usepackage{lscape}

\usepackage[colorlinks=true,citecolor=black,linkcolor=black,urlcolor=blue]{hyperref}

\newcommand{\comments}[1]{}

\renewcommand{\phi}{\varphi}

\newcommand{\ignore}[1]{}

\begin{document}

\begin{frontmatter}



\title{The packing chromatic number of the infinite square lattice is between 13 and 15}


\author[label1]{Barnaby Martin}
\author[label2]{Franco Raimondi\corref{cor2}}
\author[label2]{Taolue Chen\corref{cor3}}
\author[label3]{Jos Martin} 

\address[label1]{School of Engineering and Computing Sciences, University of Durham, DH1 3LE, U.K.}
\address[label2]{School of Science and Technology, Middlesex University, The Burroughs, Hendon, London NW4 4BT}
\address[label3]{The MathWorks, Matrix House, Cambridge, CB4 0HH, U.K.}

\cortext[cor2]{Supported by EPSRC grant EP/K033921/1.}
\cortext[cor3]{Supported by EPSRC grant.}



\begin{abstract}
Using a SAT-solver on top of a partial previously-known solution we improve the upper bound of the packing chromatic number of the infinite square lattice from 17 to 15. We discuss the merits of SAT-solving for this kind of problem as well as compare the performance of different encodings. Further, we improve the lower bound from 12 to 13 again using a SAT-solver, demonstrating the versatility of this technology for our approach.
\end{abstract}

\begin{keyword}
Packing Chromatic Number \sep Broadcast Chromatic Number \sep Graph Coloring \sep SAT Solving
\MSC 05C15
\end{keyword}
 
\end{frontmatter}

\section{Introduction}

The notion of packing colouring comes from the area of frequency planning in wireless networks, and was introduced by Goddard et \mbox{al.} in \cite{BroadcastChromaticNumber} under the name \emph{broadcast colouring}. A \emph{packing $k$-colouring} of a graph $G$ is a partition of $V(G)$ into disjoint sets $X_1$, \ldots, $X_k$, so that, for each $i \in \{1,\ldots,k\}$ and $x,y \in X_i$, the minimum distance between $x$ and $y$ in $G$, $\mathit{d}_G(x, y)$, is greater than $i$.  In other words, vertices with the same colour $i$ are pairwise at distance greater than $i$. The \emph{packing chromatic number} of a graph $G$, denoted by $\chi_p(G)$, is the smallest integer $k$ so that there exists a packing $k$-colouring. A \emph{packing colouring} is a packing $k$-colouring, for some $k$, and we sometimes drop the descriptor ``packing'', when it is clear from the context, to talk simply of a ($k$-)colouring. Packing colourings have application in frequency planning where one might imagine a broadcast at a higher wavelength travelling further, thus retransmission towers would not be required at such proximity as those for lower wavelengths (see \cite{BroadcastChromaticNumber}). Broadcast colouring seems to have been renamed packing colouring in the work \cite{HexagonalLattice}.  

The \emph{Cartesian product} of two graphs $G$ and $H$, denoted $G \Box H$, is the graph with vertex set $G \times H$ and edge set
\[ \{((x_1,y_1),(x_2,y_2)) : (x_1=x_2 \wedge (y_1,y_2)\in E(H)) \vee (y_1=y_2 \wedge (x_1,x_2)\in E(G)) \}.\]
The infinite square lattice (grid) $P_\mathbb{Z} \Box P_\mathbb{Z}$ is the graph with vertex set $\mathbb{Z} \times \mathbb{Z}$ and edge set 
\[ \{((x_1,y_1),(x_2,y_2)) : (x_1=x_2 \wedge |y_1-y_2|=1) \vee  (y_1=y_2 \wedge |x_1-x_2|=1)\}.\]
If $P_\mathbb{Z}$ is the graph with vertices $\mathbb{Z}$ and edges $(x,y)$ given by $|x-y|=1$, then the infinite square lattice is the product  $P_\mathbb{Z} \Box P_\mathbb{Z}$, which explains our notation.

The packing chromatic number of the infinite square lattice,  $\chi_p(P_\mathbb{Z} \Box P_\mathbb{Z})$, has been the topic of a number of papers. Goddard et \mbox{al.} showed in \cite{BroadcastChromaticNumber} that $\chi_p(P_\mathbb{Z} \Box P_\mathbb{Z})$ is finite, more precisely between $9$ and $23$ (inclusive). In contrast, the packing chromatic number of the infinite triangular lattice is infinite \cite{FinbowRall07}, though the packing chromatic number of the infinite hexagonal lattice is $7$ \cite{HexagonalLattice7}. The upper bound of \cite{BroadcastChromaticNumber} is witnessed by a finite grid of dimension $m \times m$ which can be endlessly translated up-down and left-right in order to periodically cover the plane. We call this a \emph{periodic} packing colouring. Such a periodic packing colouring may be seen as a packing colouring of the product $C_m \Box C_m$, where $C_m$ is the undirected $m$-cycle.

Fiala and Lidick\'y \cite{FialaLidicky07} then improved the lower bound to $10$, and Schwenk \cite{Schwenk02} improved the upper bound to $22$. Later, Ekstein, Fiala, Holub and Lidick\'y used a computer to improve the lower bound to 12 \cite{EksteinFialaHolubLidicky07} and Soukal and Holub used a clever Simulated Annealing algorithm to improve the upper bound to $17$ \cite{SoukalHolub10}. Thus these last bounds, in contrast to those that went before, both made fundamental use of mechanical computation.

\vspace{0.2cm}
\noindent \textbf{Further related work}. We are not the first to use SAT-solvers in Discrete Mathematics, especially at the interface between Combinatorics and Number Theory. Fascinating progress has been made towards the computation of van der Waerden \cite{VanDerWaerden,NewVanDerWaerden} and Ramsey \cite{DBLP:journals/corr/abs-1212-1328} numbers (see also the thesis \cite{KugeleThesis}). Indeed, the case $c=2$ of the Erd\"os Discrepancy Conjecture has been settled using this technology \cite{Konev2014}. Furthermore, we not the first to use SAT-solving techniques in packing colouring \cite{ShaoVesel} (though we were not aware of this article when we obtained our results). In \cite{ShaoVesel}, the authors translate questions of packing colouring for various finite graphs (including grids) to SAT problems and instances of Integer Programming. The paper \cite{OtherShaoVesel}, by the same authors, which relates similar techniques for other graph colouring problems, should also be mentioned here. 

We note a recent contribution on the subject of packing chromatic numbers of infinite lattices from France \cite{GastineauKT15}. In this paper, packing distances considered for the colours may be specified individually (whereas for us the packing distance and the colour number coincide). A number of works, both older and new, address questions of computational complexity for determining packing chromatic numbers. It is known that determing the packing chromatic number for general graphs is NP-hard \cite{BroadcastChromaticNumber}; indeed this remains NP-hard even for trees \cite{FialaG10}! Regarding packing chromatic number for Cartesian products of cycles, we should additionally mention the work of \cite{Jacobs2013}.

\vspace{0.2cm}
\noindent \textbf{Our story}. 
The first author heard of this problem at a talk by Bernard Lidick\'y at the 8th Slovenian Conference on Graph Theory (Bled 2011). While this problem may not be especially important, few who worked on it can doubt that it is very addictive, and further provides a vehicle through which to ponder different algorithmic techniques. One of the curiosities of the problem is that we have little theoretical insight into it. For example, suppose there is a packing colouring for the infinite square lattice involving $k$ colours:
\begin{itemize}
\item does there exist an $m$ together with an $m \times m$ grid that witnesses a periodic packing colouring with $k$?
\item does there exist a packing colouring with $k$ that has colour $1$ at maximal density ($1/2$) asymptotically?
\item does there exist a packing colouring with $k$ so that, for $i<j\leq k$ the asymptotic frequency of colour $i$ is no more than the asymptotic frequency of $j$?
\end{itemize}
The answers to the above questions are still not known. Note that it is not possible to cover asymptotically more than half of the vertices of the infinite square lattice with the colour $1$.

\vspace{0.2cm}
\noindent \textbf{Our contribution}. 
In the present note we improve the upper bound from $17$ to $16$ and then to $15$. As with all these upper bounds we give a packing colouring based on a finite grid which can be translated up-down and left-right to give a periodic packing colouring of the infinite square lattice (grid). We make essential use of the periodic $24 \times 24$ $17$-colouring given by Soukal-Holub in \cite{SoukalHolub10}, which is drawn in Figure~\ref{fig:17-col}. 

\begin{figure}[h]
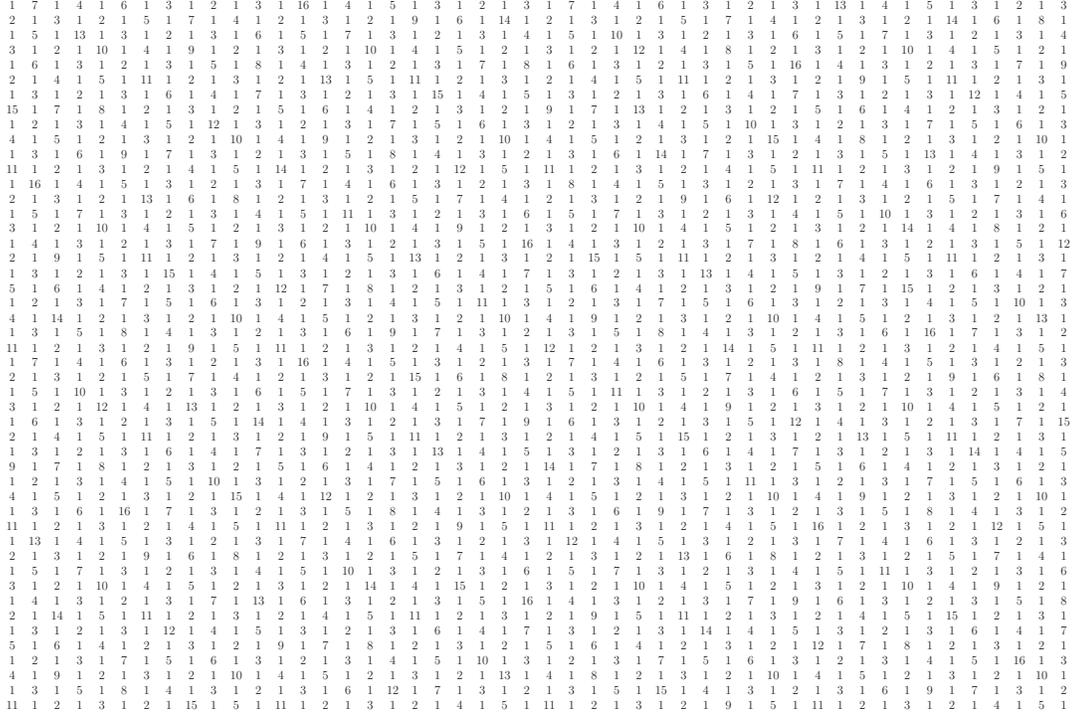

\begin{center}
\resizebox{!}{4.8cm}{
\ensuremath{
\mathbb{
\begin{array}{cccccccccccccccccccccccccccccccccccccccccccccccc}
 1 & 7 & 1 & 4 & 1 & 6 & 1 & 3 & 1 & 2 & 1 & 3 & 1 &16 & 1 & 4 & 1 & 5 & 1 & 3 & 1 & 2 & 1 & 3 & 1 & 7 & 1 & 4 & 1 & 6 & 1 & 3 & 1 & 2 & 1 & 3 & 1 &13 & 1 & 4 & 1 & 5 & 1 & 3 & 1 & 2 & 1 & 3 \\ 
 2 & 1 & 3 & 1 & 2 & 1 & 5 & 1 & 7 & 1 & 4 & 1 & 2 & 1 & 3 & 1 & 2 & 1 & 9 & 1 & 6 & 1 &14 & 1 & 2 & 1 & 3 & 1 & 2 & 1 & 5 & 1 & 7 & 1 & 4 & 1 & 2 & 1 & 3 & 1 & 2 & 1 &14 & 1 & 6 & 1 & 8 & 1 \\ 
 1 & 5 & 1 &13 & 1 & 3 & 1 & 2 & 1 & 3 & 1 & 6 & 1 & 5 & 1 & 7 & 1 & 3 & 1 & 2 & 1 & 3 & 1 & 4 & 1 & 5 & 1 &10 & 1 & 3 & 1 & 2 & 1 & 3 & 1 & 6 & 1 & 5 & 1 & 7 & 1 & 3 & 1 & 2 & 1 & 3 & 1 & 4 \\ 
 3 & 1 & 2 & 1 &10 & 1 & 4 & 1 & 9 & 1 & 2 & 1 & 3 & 1 & 2 & 1 &10 & 1 & 4 & 1 & 5 & 1 & 2 & 1 & 3 & 1 & 2 & 1 &12 & 1 & 4 & 1 & 8 & 1 & 2 & 1 & 3 & 1 & 2 & 1 &10 & 1 & 4 & 1 & 5 & 1 & 2 & 1 \\ 
 1 & 6 & 1 & 3 & 1 & 2 & 1 & 3 & 1 & 5 & 1 & 8 & 1 & 4 & 1 & 3 & 1 & 2 & 1 & 3 & 1 & 7 & 1 & 8 & 1 & 6 & 1 & 3 & 1 & 2 & 1 & 3 & 1 & 5 & 1 &16 & 1 & 4 & 1 & 3 & 1 & 2 & 1 & 3 & 1 & 7 & 1 & 9 \\ 
 2 & 1 & 4 & 1 & 5 & 1 &11 & 1 & 2 & 1 & 3 & 1 & 2 & 1 &13 & 1 & 5 & 1 &11 & 1 & 2 & 1 & 3 & 1 & 2 & 1 & 4 & 1 & 5 & 1 &11 & 1 & 2 & 1 & 3 & 1 & 2 & 1 & 9 & 1 & 5 & 1 &11 & 1 & 2 & 1 & 3 & 1\\ 
 1 & 3 & 1 & 2 & 1 & 3 & 1 & 6 & 1 & 4 & 1 & 7 & 1 & 3 & 1 & 2 & 1 & 3 & 1 &15 & 1 & 4 & 1 & 5 & 1 & 3 & 1 & 2 & 1 & 3 & 1 & 6 & 1 & 4 & 1 & 7 & 1 & 3 & 1 & 2 & 1 & 3 & 1 &12 & 1 & 4 & 1 & 5 \\ 
15 & 1 & 7 & 1 & 8 & 1 & 2 & 1 & 3 & 1 & 2 & 1 & 5 & 1 & 6 & 1 & 4 & 1 & 2 & 1 & 3 & 1 & 2 & 1 & 9 & 1 & 7 & 1 &13 & 1 & 2 & 1 & 3 & 1 & 2 & 1 & 5 & 1 & 6 & 1 & 4 & 1 & 2 & 1 & 3 & 1 & 2 & 1 \\ 
 1 & 2 & 1 & 3 & 1 & 4 & 1 & 5 & 1 &12 & 1 & 3 & 1 & 2 & 1 & 3 & 1 & 7 & 1 & 5 & 1 & 6 & 1 & 3 & 1 & 2 & 1 & 3 & 1 & 4 & 1 & 5 & 1 &10 & 1 & 3 & 1 & 2 & 1 & 3 & 1 & 7 & 1 & 5 & 1 & 6 & 1 & 3 \\ 
 4 & 1 & 5 & 1 & 2 & 1 & 3 & 1 & 2 & 1 &10 & 1 & 4 & 1 & 9 & 1 & 2 & 1 & 3 & 1 & 2 & 1 &10 & 1 & 4 & 1 & 5 & 1 & 2 & 1 & 3 & 1 & 2 & 1 &15 & 1 & 4 & 1 & 8 & 1 & 2 & 1 & 3 & 1 & 2 & 1 &10 & 1 \\ 
 1 & 3 & 1 & 6 & 1 & 9 & 1 & 7 & 1 & 3 & 1 & 2 & 1 & 3 & 1 & 5 & 1 & 8 & 1 & 4 & 1 & 3 & 1 & 2 & 1 & 3 & 1 & 6 & 1 &14 & 1 & 7 & 1 & 3 & 1 & 2 & 1 & 3 & 1 & 5 & 1 &13 & 1 & 4 & 1 & 3 & 1 & 2 \\ 
11 & 1 & 2 & 1 & 3 & 1 & 2 & 1 & 4 & 1 & 5 & 1 &14 & 1 & 2 & 1 & 3 & 1 & 2 & 1 &12 & 1 & 5 & 1 &11 & 1 & 2 & 1 & 3 & 1 & 2 & 1 & 4 & 1 & 5 & 1 &11 & 1 & 2 & 1 & 3 & 1 & 2 & 1 & 9 & 1 & 5 & 1 \\ 
 1 &16 & 1 & 4 & 1 & 5 & 1 & 3 & 1 & 2 & 1 & 3 & 1 & 7 & 1 & 4 & 1 & 6 & 1 & 3 & 1 & 2 & 1 & 3 & 1 & 8 & 1 & 4 & 1 & 5 & 1 & 3 & 1 & 2 & 1 & 3 & 1 & 7 & 1 & 4 & 1 & 6 & 1 & 3 & 1 & 2 & 1 & 3 \\ 
 2 & 1 & 3 & 1 & 2 & 1 &13 & 1 & 6 & 1 & 8 & 1 & 2 & 1 & 3 & 1 & 2 & 1 & 5 & 1 & 7 & 1 & 4 & 1 & 2 & 1 & 3 & 1 & 2 & 1 & 9 & 1 & 6 & 1 &12 & 1 & 2 & 1 & 3 & 1 & 2 & 1 & 5 & 1 & 7 & 1 & 4 & 1 \\ 
 1 & 5 & 1 & 7 & 1 & 3 & 1 & 2 & 1 & 3 & 1 & 4 & 1 & 5 & 1 &11 & 1 & 3 & 1 & 2 & 1 & 3 & 1 & 6 & 1 & 5 & 1 & 7 & 1 & 3 & 1 & 2 & 1 & 3 & 1 & 4 & 1 & 5 & 1 &10 & 1 & 3 & 1 & 2 & 1 & 3 & 1 & 6 \\ 
 3 & 1 & 2 & 1 &10 & 1 & 4 & 1 & 5 & 1 & 2 & 1 & 3 & 1 & 2 & 1 &10 & 1 & 4 & 1 & 9 & 1 & 2 & 1 & 3 & 1 & 2 & 1 &10 & 1 & 4 & 1 & 5 & 1 & 2 & 1 & 3 & 1 & 2 & 1 &14 & 1 & 4 & 1 & 8 & 1 & 2 & 1 \\ 
 1 & 4 & 1 & 3 & 1 & 2 & 1 & 3 & 1 & 7 & 1 & 9 & 1 & 6 & 1 & 3 & 1 & 2 & 1 & 3 & 1 & 5 & 1 &16 & 1 & 4 & 1 & 3 & 1 & 2 & 1 & 3 & 1 & 7 & 1 & 8 & 1 & 6 & 1 & 3 & 1 & 2 & 1 & 3 & 1 & 5 & 1 &12 \\ 
 2 & 1 & 9 & 1 & 5 & 1 &11 & 1 & 2 & 1 & 3 & 1 & 2 & 1 & 4 & 1 & 5 & 1 &13 & 1 & 2 & 1 & 3 & 1 & 2 & 1 &15 & 1 & 5 & 1 &11 & 1 & 2 & 1 & 3 & 1 & 2 & 1 & 4 & 1 & 5 & 1 &11 & 1 & 2 & 1 & 3 & 1 \\ 
 1 & 3 & 1 & 2 & 1 & 3 & 1 &15 & 1 & 4 & 1 & 5 & 1 & 3 & 1 & 2 & 1 & 3 & 1 & 6 & 1 & 4 & 1 & 7 & 1 & 3 & 1 & 2 & 1 & 3 & 1 &13 & 1 & 4 & 1 & 5 & 1 & 3 & 1 & 2 & 1 & 3 & 1 & 6 & 1 & 4 & 1 & 7 \\ 
 5 & 1 & 6 & 1 & 4 & 1 & 2 & 1 & 3 & 1 & 2 & 1 &12 & 1 & 7 & 1 & 8 & 1 & 2 & 1 & 3 & 1 & 2 & 1 & 5 & 1 & 6 & 1 & 4 & 1 & 2 & 1 & 3 & 1 & 2 & 1 & 9 & 1 & 7 & 1 &15 & 1 & 2 & 1 & 3 & 1 & 2 & 1 \\ 
 1 & 2 & 1 & 3 & 1 & 7 & 1 & 5 & 1 & 6 & 1 & 3 & 1 & 2 & 1 & 3 & 1 & 4 & 1 & 5 & 1 &11 & 1 & 3 & 1 & 2 & 1 & 3 & 1 & 7 & 1 & 5 & 1 & 6 & 1 & 3 & 1 & 2 & 1 & 3 & 1 & 4 & 1 & 5 & 1 &10 & 1 & 3 \\ 
 4 & 1 &14 & 1 & 2 & 1 & 3 & 1 & 2 & 1 &10 & 1 & 4 & 1 & 5 & 1 & 2 & 1 & 3 & 1 & 2 & 1 &10 & 1 & 4 & 1 & 9 & 1 & 2 & 1 & 3 & 1 & 2 & 1 &10 & 1 & 4 & 1 & 5 & 1 & 2 & 1 & 3 & 1 & 2 & 1 &13 & 1 \\ 
 1 & 3 & 1 & 5 & 1 & 8 & 1 & 4 & 1 & 3 & 1 & 2 & 1 & 3 & 1 & 6 & 1 & 9 & 1 & 7 & 1 & 3 & 1 & 2 & 1 & 3 & 1 & 5 & 1 & 8 & 1 & 4 & 1 & 3 & 1 & 2 & 1 & 3 & 1 & 6 & 1 &16 & 1 & 7 & 1 & 3 & 1 & 2 \\ 
11 & 1 & 2 & 1 & 3 & 1 & 2 & 1 & 9 & 1 & 5 & 1 &11 & 1 & 2 & 1 & 3 & 1 & 2 & 1 & 4 & 1 & 5 & 1 &12 & 1 & 2 & 1 & 3 & 1 & 2 & 1 &14 & 1 & 5 & 1 &11 & 1 & 2 & 1 & 3 & 1 & 2 & 1 & 4 & 1 & 5 & 1 \\ 
 1 & 7 & 1 & 4 & 1 & 6 & 1 & 3 & 1 & 2 & 1 & 3 & 1 &16 & 1 & 4 & 1 & 5 & 1 & 3 & 1 & 2 & 1 & 3 & 1 & 7 & 1 & 4 & 1 & 6 & 1 & 3 & 1 & 2 & 1 & 3 & 1 & 8 & 1 & 4 & 1 & 5 & 1 & 3 & 1 & 2 & 1 & 3 \\ 
 2 & 1 & 3 & 1 & 2 & 1 & 5 & 1 & 7 & 1 & 4 & 1 & 2 & 1 & 3 & 1 & 2 & 1 &15 & 1 & 6 & 1 & 8 & 1 & 2 & 1 & 3 & 1 & 2 & 1 & 5 & 1 & 7 & 1 & 4 & 1 & 2 & 1 & 3 & 1 & 2 & 1 & 9 & 1 & 6 & 1 & 8 & 1 \\ 
 1 & 5 & 1 &10 & 1 & 3 & 1 & 2 & 1 & 3 & 1 & 6 & 1 & 5 & 1 & 7 & 1 & 3 & 1 & 2 & 1 & 3 & 1 & 4 & 1 & 5 & 1 &11 & 1 & 3 & 1 & 2 & 1 & 3 & 1 & 6 & 1 & 5 & 1 & 7 & 1 & 3 & 1 & 2 & 1 & 3 & 1 & 4 \\ 
 3 & 1 & 2 & 1 &12 & 1 & 4 & 1 &13 & 1 & 2 & 1 & 3 & 1 & 2 & 1 &10 & 1 & 4 & 1 & 5 & 1 & 2 & 1 & 3 & 1 & 2 & 1 &10 & 1 & 4 & 1 & 9 & 1 & 2 & 1 & 3 & 1 & 2 & 1 &10 & 1 & 4 & 1 & 5 & 1 & 2 & 1 \\ 
 1 & 6 & 1 & 3 & 1 & 2 & 1 & 3 & 1 & 5 & 1 &14 & 1 & 4 & 1 & 3 & 1 & 2 & 1 & 3 & 1 & 7 & 1 & 9 & 1 & 6 & 1 & 3 & 1 & 2 & 1 & 3 & 1 & 5 & 1 &12 & 1 & 4 & 1 & 3 & 1 & 2 & 1 & 3 & 1 & 7 & 1 &15 \\ 
 2 & 1 & 4 & 1 & 5 & 1 &11 & 1 & 2 & 1 & 3 & 1 & 2 & 1 & 9 & 1 & 5 & 1 &11 & 1 & 2 & 1 & 3 & 1 & 2 & 1 & 4 & 1 & 5 & 1 &15 & 1 & 2 & 1 & 3 & 1 & 2 & 1 &13 & 1 & 5 & 1 &11 & 1 & 2 & 1 & 3 & 1 \\ 
 1 & 3 & 1 & 2 & 1 & 3 & 1 & 6 & 1 & 4 & 1 & 7 & 1 & 3 & 1 & 2 & 1 & 3 & 1 &13 & 1 & 4 & 1 & 5 & 1 & 3 & 1 & 2 & 1 & 3 & 1 & 6 & 1 & 4 & 1 & 7 & 1 & 3 & 1 & 2 & 1 & 3 & 1 &14 & 1 & 4 & 1 & 5 \\ 
 9 & 1 & 7 & 1 & 8 & 1 & 2 & 1 & 3 & 1 & 2 & 1 & 5 & 1 & 6 & 1 & 4 & 1 & 2 & 1 & 3 & 1 & 2 & 1 &14 & 1 & 7 & 1 & 8 & 1 & 2 & 1 & 3 & 1 & 2 & 1 & 5 & 1 & 6 & 1 & 4 & 1 & 2 & 1 & 3 & 1 & 2 & 1 \\ 
 1 & 2 & 1 & 3 & 1 & 4 & 1 & 5 & 1 &10 & 1 & 3 & 1 & 2 & 1 & 3 & 1 & 7 & 1 & 5 & 1 & 6 & 1 & 3 & 1 & 2 & 1 & 3 & 1 & 4 & 1 & 5 & 1 &11 & 1 & 3 & 1 & 2 & 1 & 3 & 1 & 7 & 1 & 5 & 1 & 6 & 1 & 3 \\ 
 4 & 1 & 5 & 1 & 2 & 1 & 3 & 1 & 2 & 1 &15 & 1 & 4 & 1 &12 & 1 & 2 & 1 & 3 & 1 & 2 & 1 &10 & 1 & 4 & 1 & 5 & 1 & 2 & 1 & 3 & 1 & 2 & 1 &10 & 1 & 4 & 1 & 9 & 1 & 2 & 1 & 3 & 1 & 2 & 1 &10 & 1 \\ 
 1 & 3 & 1 & 6 & 1 &16 & 1 & 7 & 1 & 3 & 1 & 2 & 1 & 3 & 1 & 5 & 1 & 8 & 1 & 4 & 1 & 3 & 1 & 2 & 1 & 3 & 1 & 6 & 1 & 9 & 1 & 7 & 1 & 3 & 1 & 2 & 1 & 3 & 1 & 5 & 1 & 8 & 1 & 4 & 1 & 3 & 1 & 2 \\ 
11 & 1 & 2 & 1 & 3 & 1 & 2 & 1 & 4 & 1 & 5 & 1 &11 & 1 & 2 & 1 & 3 & 1 & 2 & 1 & 9 & 1 & 5 & 1 &11 & 1 & 2 & 1 & 3 & 1 & 2 & 1 & 4 & 1 & 5 & 1 &16 & 1 & 2 & 1 & 3 & 1 & 2 & 1 &12 & 1 & 5 & 1 \\ 
 1 &13 & 1 & 4 & 1 & 5 & 1 & 3 & 1 & 2 & 1 & 3 & 1 & 7 & 1 & 4 & 1 & 6 & 1 & 3 & 1 & 2 & 1 & 3 & 1 &12 & 1 & 4 & 1 & 5 & 1 & 3 & 1 & 2 & 1 & 3 & 1 & 7 & 1 & 4 & 1 & 6 & 1 & 3 & 1 & 2 & 1 & 3 \\ 
 2 & 1 & 3 & 1 & 2 & 1 & 9 & 1 & 6 & 1 & 8 & 1 & 2 & 1 & 3 & 1 & 2 & 1 & 5 & 1 & 7 & 1 & 4 & 1 & 2 & 1 & 3 & 1 & 2 & 1 &13 & 1 & 6 & 1 & 8 & 1 & 2 & 1 & 3 & 1 & 2 & 1 & 5 & 1 & 7 & 1 & 4 & 1 \\ 
 1 & 5 & 1 & 7 & 1 & 3 & 1 & 2 & 1 & 3 & 1 & 4 & 1 & 5 & 1 &10 & 1 & 3 & 1 & 2 & 1 & 3 & 1 & 6 & 1 & 5 & 1 & 7 & 1 & 3 & 1 & 2 & 1 & 3 & 1 & 4 & 1 & 5 & 1 &11 & 1 & 3 & 1 & 2 & 1 & 3 & 1 & 6 \\ 
 3 & 1 & 2 & 1 &10 & 1 & 4 & 1 & 5 & 1 & 2 & 1 & 3 & 1 & 2 & 1 &14 & 1 & 4 & 1 &15 & 1 & 2 & 1 & 3 & 1 & 2 & 1 &10 & 1 & 4 & 1 & 5 & 1 & 2 & 1 & 3 & 1 & 2 & 1 &10 & 1 & 4 & 1 & 9 & 1 & 2 & 1 \\ 
 1 & 4 & 1 & 3 & 1 & 2 & 1 & 3 & 1 & 7 & 1 &13 & 1 & 6 & 1 & 3 & 1 & 2 & 1 & 3 & 1 & 5 & 1 &16 & 1 & 4 & 1 & 3 & 1 & 2 & 1 & 3 & 1 & 7 & 1 & 9 & 1 & 6 & 1 & 3 & 1 & 2 & 1 & 3 & 1 & 5 & 1 & 8 \\ 
 2 & 1 &14 & 1 & 5 & 1 &11 & 1 & 2 & 1 & 3 & 1 & 2 & 1 & 4 & 1 & 5 & 1 &11 & 1 & 2 & 1 & 3 & 1 & 2 & 1 & 9 & 1 & 5 & 1 &11 & 1 & 2 & 1 & 3 & 1 & 2 & 1 & 4 & 1 & 5 & 1 &15 & 1 & 2 & 1 & 3 & 1 \\ 
 1 & 3 & 1 & 2 & 1 & 3 & 1 &12 & 1 & 4 & 1 & 5 & 1 & 3 & 1 & 2 & 1 & 3 & 1 & 6 & 1 & 4 & 1 & 7 & 1 & 3 & 1 & 2 & 1 & 3 & 1 &14 & 1 & 4 & 1 & 5 & 1 & 3 & 1 & 2 & 1 & 3 & 1 & 6 & 1 & 4 & 1 & 7 \\ 
 5 & 1 & 6 & 1 & 4 & 1 & 2 & 1 & 3 & 1 & 2 & 1 & 9 & 1 & 7 & 1 & 8 & 1 & 2 & 1 & 3 & 1 & 2 & 1 & 5 & 1 & 6 & 1 & 4 & 1 & 2 & 1 & 3 & 1 & 2 & 1 &12 & 1 & 7 & 1 & 8 & 1 & 2 & 1 & 3 & 1 & 2 & 1 \\ 
 1 & 2 & 1 & 3 & 1 & 7 & 1 & 5 & 1 & 6 & 1 & 3 & 1 & 2 & 1 & 3 & 1 & 4 & 1 & 5 & 1 &10 & 1 & 3 & 1 & 2 & 1 & 3 & 1 & 7 & 1 & 5 & 1 & 6 & 1 & 3 & 1 & 2 & 1 & 3 & 1 & 4 & 1 & 5 & 1 &16 & 1 & 3 \\ 
 4 & 1 & 9 & 1 & 2 & 1 & 3 & 1 & 2 & 1 &10 & 1 & 4 & 1 & 5 & 1 & 2 & 1 & 3 & 1 & 2 & 1 &13 & 1 & 4 & 1 & 8 & 1 & 2 & 1 & 3 & 1 & 2 & 1 &10 & 1 & 4 & 1 & 5 & 1 & 2 & 1 & 3 & 1 & 2 & 1 &10 & 1 \\ 
 1 & 3 & 1 & 5 & 1 & 8 & 1 & 4 & 1 & 3 & 1 & 2 & 1 & 3 & 1 & 6 & 1 &12 & 1 & 7 & 1 & 3 & 1 & 2 & 1 & 3 & 1 & 5 & 1 &15 & 1 & 4 & 1 & 3 & 1 & 2 & 1 & 3 & 1 & 6 & 1 & 9 & 1 & 7 & 1 & 3 & 1 & 2 \\ 
11 & 1 & 2 & 1 & 3 & 1 & 2 & 1 &15 & 1 & 5 & 1 &11 & 1 & 2 & 1 & 3 & 1 & 2 & 1 & 4 & 1 & 5 & 1 &11 & 1 & 2 & 1 & 3 & 1 & 2 & 1 & 9 & 1 & 5 & 1 &11 & 1 & 2 & 1 & 3 & 1 & 2 & 1 & 4 & 1 & 5 & 1 \\
\end{array}
}
}
}
\end{center}
\label{fig:main}
\caption{Our $48 \times 48$ periodic solution in $16$ colours}
\end{figure}

For our $16$-colouring, we take the Soukal-Hulob colouring and remove colours $8$ to $17$, then we blow this up from $24 \times 24$ to $48 \times 48$ by taking four copies of it ($2 \times 2$ in shape). We then give the resulting partially coloured grid to a SAT-solver to see if a $16$-colouring is possible, which it turns out it is. Our $16$-colouring is specified as the obvious periodic translation of the colouring in Figure~\ref{fig:main}. Note that our method with a SAT-solver does not run efficiently unless several colours are \emph{planted}, that is some entries (vertices) are preset to a certain colour. For example, with just colour $1$ planted at maximal density of $1/2$ on a $24 \times 24$ grid, the computation runs, on the machines we are using, indefinitely. Planting colours may be seen as a form of \emph{pre-colouring}, thus the objects given to the SAT-solver are partially (pre-)coloured grids. Planting colours allows us also to break the problem's symmetries.

For our $15$-colouring, we take the Soukal-Hulob colouring and remove colours $6$ to $17$, then blow this up from $24 \times 24$ to $72 \times 72$. This colouring is depicted in Figure~\ref{fig:15}.

We have additionally improved the lower bound by ruling out the possibility of a certain $12$-colouring on a $14 \times 14$ grid. Here we derive our result by producing a SAT instance that is found to be unsatisfiable.

\section{Encoding and computation}

Let $[n]:=\{1,\ldots,n\}$. 
Our \emph{basic encoding}, for investigating the upper bound, involves variables $P_{i,j,k}$, whose being true asserts that position $(i,j)$ in some $m$-colouring, periodic on a grid of size $n \times n$, is set to colour $k$. We thus need big clauses of the form $P_{i,j,1} \vee P_{i,j,2} \vee \ldots \vee P_{i,j,m}$ for each $(i,j) \in [n]^2$, together with constraints $\neg P_{i,j,k} \vee \neg P_{i',j',k}$ whenever the distance between $(i,j)$ and $(i',j')$, $d((i,j),(i',j'))$, is less than $k$. This distance must, of course be calculated toroidally, \mbox{i.e.}  $d((i,j),(i',j'))=\min(i-i' \bmod n, i'-i \bmod n)+\min(j-j' \bmod n, j'-j \bmod n)$. To plant colour $k$ at position $(i,j)$ we use the singleton clause $P_{i,j,k}$.

Our \emph{commander encoding} (see \cite{Commander}) has the same pair clauses as the basic encoding, but splits the big clauses into a system of smaller ones. In considering a $16$-colouring, each clause $P_{i,j,1} \vee P_{i,j,2} \vee \ldots \vee P_{i,j,16}$ is substituted by five clauses, requiring four new commander variables $C_{i,j,1},\ldots,C_{i,j,4}$, as follows.
\[
\begin{array}{cc}
\neg C_{i,j,1} \vee  P_{i,j,1} \vee P_{i,j,2} \vee P_{i,j,3} \vee P_{i,j,4} \\
\neg C_{i,j,2} \vee  P_{i,j,5} \vee P_{i,j,6} \vee P_{i,j,7} \vee P_{i,j,8} \\
\neg C_{i,j,3} \vee  P_{i,j,9} \vee P_{i,j,10} \vee P_{i,j,11} \vee P_{i,j,12} \\
\neg C_{i,j,4} \vee  P_{i,j,13} \vee P_{i,j,14} \vee P_{i,j,15} \vee P_{i,j,16} \\
C_{i,j,1} \vee  C_{i,j,2} \vee C_{i,j,3} \vee C_{i,j,4} \\
\end{array}
\]
Clearly, this system is equivalent to the single big clause, with satisfying assignments in the variables $P_{i,j,k}$ being preserved. For $15$-colourings, each big clause is substituted by six clauses involving five new commander variables (thus five of those clauses have width $4$, and the fifth, $5$) in the analogous fashion. 

We used the SAT-solver Glucose Syrup on a quad core Intel(R) Xeon(R) CPU E5-2660 0 @ 2.20GHz with 40 GB RAM, running Ubuntu Linux 14.04. In all the experiments we set a maximum limit of 12 GB of RAM, which allowed four threads of the SAT-solver to run in parallel (no hyperthreading, just one process per physical CPU, and no swapping).

Glucose Syrup gave a wide variance in its results so in the table in Figure~\ref{fig:running-times} each result is the average over five runs (three runs for the starred rows whose times are longer).

\subsection{The lower bound}

In a finite $m \times n$ grid, we refer to the entry at row $i$ and column $j$ by $(i,j)$. The lower bound from \cite{EksteinFialaHolubLidicky07} considers colouring a $15 \times 9$ rectangle in $11$ colours but forgetting the toroidal constraints (\mbox{i.e.} colour $11$ might appear legally in the middle of the first row as well as in the middle of the last row, which would obviously be forbidden in a periodic colouring). Indeed, the computation proves that no such $11$-colouring exists (on the innocuous assumption, w.l.o.g, that colour $9$ is placed in position $(5,5)$). Their computation, admittedly in 2009, took $120$ days. We have replicated their lower bound with our SAT-solver, on a $12 \times 12$ grid, with $9$ planted in position $(6,6)$, in just 26952 seconds. 

Finally we have improved the lower bound by showing that $14 \times 14$ with $(7,12)$ planted $9$ for a $12$-colouring is not possible. Here the unsatisfiability was found after $16$ days. This comprehensively demonstrates the versatility of the SAT-solving method for the lower bound as well as the upper bound. 

Our SAT instances are available for download from the website with address www.bedewell.com/sat. The lower bound instance, for historical reasons, bears the epithet ``sheep''.
 
\begin{figure}
\begin{center}
\resizebox{!}{4cm}{
\ensuremath{
\mathbb{
\begin{array}{cccccccccccccccccccccccc}
 1 &  7 &  1 &  4 &  1 &  6 &  1 &  3 &  1 &  2 &  1 &  3 &  1 &  8 &  1 &  4 &  1 &  5 &  1 &  3 &  1 &  2 &  1 &  3 \\ 
 2 &  1 &  3 &  1 &  2 &  1 &  5 &  1 &  7 &  1 &  4 &  1 &  2 &  1 &  3 &  1 &  2 &  1 &  10 &  1 &  6 &  1 &  11 &  1 \\ 
 1 &  5 &  1 &  9 &  1 &  3 &  1 &  2 &  1 &  3 &  1 &  6 &  1 &  5 &  1 &  7 &  1 &  3 &  1 &  2 &  1 &  3 &  1 &  4 \\ 
 3 &  1 &  2 &  1 &  15 &  1 &  4 &  1 &  10 &  1 &  2 &  1 &  3 &  1 &  2 &  1 &  17 &  1 &  4 &  1 &  5 &  1 &  2 &  1 \\ 
 1 &  6 &  1 &  3 &  1 &  2 &  1 &  3 &  1 &  5 &  1 &  13 &  1 &  4 &  1 &  3 &  1 &  2 &  1 &  3 &  1 &  7 &  1 &  8 \\ 
 2 &  1 &  4 &  1 &  5 &  1 &  16 &  1 &  2 &  1 &  3 &  1 &  2 &  1 &  11 &  1 &  5 &  1 &  9 &  1 &  2 &  1 &  3 &  1 \\ 
 1 &  3 &  1 &  2 &  1 &  3 &  1 &  6 &  1 &  4 &  1 &  7 &  1 &  3 &  1 &  2 &  1 &  3 &  1 &  12 &  1 &  4 &  1 &  5 \\ 
 10 &  1 &  7 &  1 &  11 &  1 &  2 &  1 &  3 &  1 &  2 &  1 &  5 &  1 &  6 &  1 &  4 &  1 &  2 &  1 &  3 &  1 &  2 &  1 \\ 
 1 &  2 &  1 &  3 &  1 &  4 &  1 &  5 &  1 &  8 &  1 &  3 &  1 &  2 &  1 &  3 &  1 &  7 &  1 &  5 &  1 &  6 &  1 &  3 \\ 
 4 &  1 &  5 &  1 &  2 &  1 &  3 &  1 &  2 &  1 &  9 &  1 &  4 &  1 &  10 &  1 &  2 &  1 &  3 &  1 &  2 &  1 &  13 &  1 \\ 
 1 &  3 &  1 &  6 &  1 &  14 &  1 &  7 &  1 &  3 &  1 &  2 &  1 &  3 &  1 &  5 &  1 &  8 &  1 &  4 &  1 &  3 &  1 &  2 \\ 
 9 &  1 &  2 &  1 &  3 &  1 &  2 &  1 &  4 &  1 &  5 &  1 &  15 &  1 &  2 &  1 &  3 &  1 &  2 &  1 &  11 &  1 &  5 &  1 \\ 
 1 &  8 &  1 &  4 &  1 &  5 &  1 &  3 &  1 &  2 &  1 &  3 &  1 &  7 &  1 &  4 &  1 &  6 &  1 &  3 &  1 &  2 &  1 &  3 \\ 
 2 &  1 &  3 &  1 &  2 &  1 &  10 &  1 &  6 &  1 & 11 &  1 &  2 &  1 &  3 &  1 &  2 &  1 &  5 &  1 &  7 &  1 &  4 &  1 \\ 
 1 &  5 &  1 &  7 &  1 &  3 &  1 &  2 &  1 &  3 &  1 &  4 &  1 &  5 &  1 &  9 &  1 &  3 &  1 &  2 &  1 &  3 &  1 &  6 \\ 
 3 &  1 &  2 &  1 &  17 &  1 &  4 &  1 &  5 &  1 &  2 &  1 &  3 &  1 &  2 &  1 &  14 &  1 &  4 &  1 &  10 &  1 &  2 &  1 \\ 
 1 &  4 &  1 &  3 &  1 &  2 &  1 &  3 &  1 &  7 &  1 &  8 &  1 &  6 &  1 &  3 &  1 &  2 &  1 &  3 &  1 &  5 &  1 &  12 \\ 
 2 &  1 &  11 &  1 &  5 &  1 &  13 &  1 &  2 &  1 &  3 &  1 &  2 &  1 &  4 &  1 &  5 &  1 &  16 &  1 &  2 &  1 &  3 &  1 \\ 
 1 &  3 &  1 &  2 &  1 &  3 &  1 &  9 &  1 &  4 &  1 &  5 &  1 &  3 &  1 &  2 &  1 &  3 &  1 &  6 &  1 &  4 &  1 &  7 \\ 
 5 &  1 &  6 &  1 &  4 &  1 &  2 &  1 &  3 &  1 &  2 &  1 &  10 &  1 &  7 &  1 &  11 &  1 &  2 &  1 &  3 &  1 &  2 &  1 \\ 
 1 &  2 &  1 &  3 &  1 &  7 &  1 &  5 &  1 &  6 &  1 &  3 &  1 &  2 &  1 &  3 &  1 &  4 &  1 &  5 &  1 &  8 &  1 &  3 \\ 
 4 &  1 &  10 &  1 &  2 &  1 &  3 &  1 &  2 &  1 &  12 &  1 &  4 &  1 &  5 &  1 &  2 &  1 &  3 &  1 &  2 &  1 &  9 &  1 \\ 
 1 &  3 &  1 &  5 &  1 &  8 &  1 &  4 &  1 &  3 &  1 &  2 &  1 &  3 &  1 &  6 &  1 &  15 &  1 &  7 &  1 &  3 &  1 &  2 \\ 
 14 &  1 &  2 &  1 &  3 &  1 &  2 &  1 &  11 &  1 &  5 &  1 &  9 &  1 &  2 &  1 &  3 &  1 &  2 &  1 &  4 &  1 &  5 &  1 \\ 
\end{array}
}
}
}
\end{center}
\label{fig:17-col}
\caption{The Soukal-Holub $17$-colouring on $24\times24$.}
\end{figure}

\begin{figure}
\begin{center}
\resizebox{!}{3.5cm}{
\ensuremath{
\begin{array}{cc}
1 & 288 \\
2 & 72 \\
3 & 72 \\
4 & 32 \\
5 & 32 \\
6 & 16 \\
7 & 16 \\
8 & 8 \\
9 & 8 \\
10 & 8 \\ 
11 & 8 \\
12 & 3 \\
13 & 3 \\
14 & 3 \\
15 & 3 \\
16 & 2 \\
17 & 2 \\
\hline
& 576
\end{array}
}
}
\hspace{1cm}
\resizebox{!}{3.9cm}{
\ensuremath{
\begin{array}{cc}
1 & 1152 \\
2 & 288 \\
3 & 288 \\
4 & 128 \\
5 & 128 \\
6 & 64 \\
7 & 64 \\
8 & 28 \\
9 & 32 \\
10 & 32 \\ 
11 & 31 \\
12 & 16 \\
13 & 16 \\
14 & 13 \\
15 & 14 \\
16 & 10 \\
\\
\hline
& 2304
\end{array}
}
}
\hspace{1cm}
\resizebox{!}{3.9cm}{
\ensuremath{
\begin{array}{cc}
1 & 2592 \\
2 & 648 \\
3 & 648 \\
4 & 288 \\
5 & 288 \\
6 & 144 \\
7 & 144 \\
8 & 72 \\
9 & 72 \\
10 & 72 \\ 
11 & 72 \\
12 & 36 \\
13 & 36 \\
14 & 36 \\
15 & 36 \\
\\
\\
\hline
& 5184
\end{array}
}
}
\end{center}
\caption{Frequency table for our $15$-colouring (right) on $72\times 72$, $16$-colouring on $48\times 48$ grid (middle) and for the Soukal-Holub $17$-colouring on $24\times 24$ (left).}
\label{fig:frequency}
\end{figure}


\begin{landscape}
\begin{figure}
\begin{center}
\resizebox{!}{8cm}{
\setlength{\tabcolsep}{1pt}
\tiny
\begin{tabular}{cccccccccccccccccccccccccccccccccccccccccccccccccccccccccccccccccccccccc}
1&2&1&3&1&2&1&12&1&4&1&7&1&2&1&3&1&2&1&5&1&4&1&15&1&2&1&3&1&2&1&6&1&4&1&13&1&2&1&3&1&2&1&5&1&4&1&8&1&2&1&3&1&2&1&7&1&4&1&6&1&2&1&3&1&2&1&5&1&4&1&9\\
7&1&5&1&6&1&3&1&2&1&3&1&8&1&5&1&4&1&3&1&2&1&3&1&9&1&5&1&7&1&3&1&2&1&3&1&9&1&5&1&4&1&3&1&2&1&3&1&6&1&5&1&13&1&3&1&2&1&3&1&9&1&5&1&4&1&3&1&2&1&3&1\\
1&3&1&2&1&4&1&11&1&5&1&2&1&3&1&2&1&7&1&11&1&6&1&2&1&3&1&2&1&4&1&10&1&5&1&2&1&3&1&2&1&12&1&11&1&7&1&2&1&3&1&2&1&4&1&11&1&5&1&2&1&3&1&2&1&6&1&10&1&14&1&2\\
4&1&8&1&3&1&2&1&3&1&6&1&4&1&9&1&3&1&2&1&3&1&5&1&4&1&8&1&3&1&2&1&3&1&7&1&4&1&6&1&3&1&2&1&3&1&5&1&4&1&9&1&3&1&2&1&3&1&12&1&4&1&7&1&3&1&2&1&3&1&5&1\\
1&2&1&15&1&5&1&7&1&2&1&3&1&2&1&13&1&5&1&4&1&2&1&3&1&2&1&6&1&5&1&14&1&2&1&3&1&2&1&8&1&5&1&4&1&2&1&3&1&2&1&7&1&5&1&6&1&2&1&3&1&2&1&8&1&5&1&4&1&2&1&3\\
6&1&3&1&2&1&3&1&4&1&10&1&5&1&3&1&2&1&3&1&12&1&10&1&7&1&3&1&2&1&3&1&4&1&11&1&5&1&3&1&2&1&3&1&6&1&10&1&14&1&3&1&2&1&3&1&4&1&10&1&5&1&3&1&2&1&3&1&7&1&11&1\\
1&5&1&4&1&9&1&2&1&3&1&2&1&7&1&4&1&6&1&2&1&3&1&2&1&5&1&4&1&9&1&2&1&3&1&2&1&15&1&4&1&7&1&2&1&3&1&2&1&5&1&4&1&15&1&2&1&3&1&2&1&6&1&4&1&9&1&2&1&3&1&2\\
3&1&2&1&3&1&6&1&5&1&14&1&3&1&2&1&3&1&8&1&5&1&4&1&3&1&2&1&3&1&7&1&5&1&6&1&3&1&2&1&3&1&9&1&5&1&4&1&3&1&2&1&3&1&8&1&5&1&7&1&3&1&2&1&3&1&13&1&5&1&4&1\\
1&10&1&7&1&2&1&3&1&2&1&4&1&11&1&5&1&2&1&3&1&2&1&6&1&11&1&13&1&2&1&3&1&2&1&4&1&10&1&5&1&2&1&3&1&2&1&7&1&11&1&6&1&2&1&3&1&2&1&4&1&11&1&5&1&2&1&3&1&2&1&12\\
2&1&3&1&5&1&4&1&8&1&3&1&2&1&3&1&15&1&4&1&7&1&3&1&2&1&3&1&5&1&4&1&8&1&3&1&2&1&3&1&6&1&4&1&8&1&3&1&2&1&3&1&5&1&4&1&9&1&3&1&2&1&3&1&7&1&4&1&6&1&3&1\\
1&4&1&2&1&3&1&2&1&7&1&5&1&6&1&2&1&3&1&2&1&9&1&5&1&4&1&2&1&3&1&2&1&12&1&5&1&7&1&2&1&3&1&2&1&13&1&5&1&4&1&2&1&3&1&2&1&6&1&5&1&14&1&2&1&3&1&2&1&8&1&5\\
3&1&6&1&11&1&13&1&3&1&2&1&3&1&4&1&10&1&5&1&3&1&2&1&3&1&7&1&10&1&6&1&3&1&2&1&3&1&4&1&11&1&5&1&3&1&2&1&3&1&12&1&10&1&7&1&3&1&2&1&3&1&4&1&10&1&5&1&3&1&2&1\\
1&2&1&3&1&2&1&5&1&4&1&9&1&2&1&3&1&2&1&6&1&4&1&14&1&2&1&3&1&2&1&5&1&4&1&9&1&2&1&3&1&2&1&7&1&4&1&6&1&2&1&3&1&2&1&5&1&4&1&8&1&2&1&3&1&2&1&15&1&4&1&7\\
9&1&5&1&4&1&3&1&2&1&3&1&12&1&5&1&7&1&3&1&2&1&3&1&8&1&5&1&4&1&3&1&2&1&3&1&6&1&5&1&14&1&3&1&2&1&3&1&9&1&5&1&4&1&3&1&2&1&3&1&7&1&5&1&6&1&3&1&2&1&3&1\\
1&3&1&2&1&7&1&10&1&6&1&2&1&3&1&2&1&4&1&11&1&5&1&2&1&3&1&2&1&15&1&11&1&7&1&2&1&3&1&2&1&4&1&10&1&5&1&2&1&3&1&2&1&6&1&11&1&13&1&2&1&3&1&2&1&4&1&11&1&5&1&2\\
4&1&14&1&3&1&2&1&3&1&5&1&4&1&8&1&3&1&2&1&3&1&7&1&4&1&6&1&3&1&2&1&3&1&5&1&4&1&8&1&3&1&2&1&3&1&15&1&4&1&7&1&3&1&2&1&3&1&5&1&4&1&12&1&3&1&2&1&3&1&6&1\\
1&2&1&8&1&5&1&4&1&2&1&3&1&2&1&6&1&5&1&13&1&2&1&3&1&2&1&9&1&5&1&4&1&2&1&3&1&2&1&7&1&5&1&6&1&2&1&3&1&2&1&8&1&5&1&4&1&2&1&3&1&2&1&9&1&5&1&7&1&2&1&3\\
5&1&3&1&2&1&3&1&15&1&11&1&7&1&3&1&2&1&3&1&4&1&10&1&5&1&3&1&2&1&3&1&6&1&10&1&13&1&3&1&2&1&3&1&4&1&11&1&5&1&3&1&2&1&3&1&7&1&10&1&6&1&3&1&2&1&3&1&4&1&10&1\\
1&7&1&4&1&6&1&2&1&3&1&2&1&5&1&4&1&9&1&2&1&3&1&2&1&12&1&4&1&7&1&2&1&3&1&2&1&5&1&4&1&9&1&2&1&3&1&2&1&6&1&4&1&14&1&2&1&3&1&2&1&5&1&4&1&8&1&2&1&3&1&2\\
3&1&2&1&3&1&9&1&5&1&4&1&3&1&2&1&3&1&7&1&5&1&6&1&3&1&2&1&3&1&8&1&5&1&4&1&3&1&2&1&3&1&12&1&5&1&7&1&3&1&2&1&3&1&9&1&5&1&4&1&3&1&2&1&3&1&6&1&5&1&13&1\\
1&11&1&5&1&2&1&3&1&2&1&6&1&10&1&14&1&2&1&3&1&2&1&4&1&11&1&5&1&2&1&3&1&2&1&7&1&11&1&6&1&2&1&3&1&2&1&4&1&10&1&5&1&2&1&3&1&2&1&15&1&11&1&7&1&2&1&3&1&2&1&4\\
2&1&3&1&12&1&4&1&7&1&3&1&2&1&3&1&5&1&4&1&8&1&3&1&2&1&3&1&6&1&4&1&14&1&3&1&2&1&3&1&5&1&4&1&8&1&3&1&2&1&3&1&7&1&4&1&6&1&3&1&2&1&3&1&5&1&4&1&9&1&3&1\\
1&6&1&2&1&3&1&2&1&8&1&5&1&4&1&2&1&3&1&2&1&15&1&5&1&7&1&2&1&3&1&2&1&9&1&5&1&4&1&2&1&3&1&2&1&6&1&5&1&13&1&2&1&3&1&2&1&8&1&5&1&4&1&2&1&3&1&2&1&7&1&5\\
3&1&4&1&10&1&5&1&3&1&2&1&3&1&7&1&11&1&6&1&3&1&2&1&3&1&4&1&10&1&5&1&3&1&2&1&3&1&15&1&10&1&7&1&3&1&2&1&3&1&4&1&11&1&5&1&3&1&2&1&3&1&6&1&10&1&14&1&3&1&2&1\\
1&2&1&3&1&2&1&6&1&4&1&13&1&2&1&3&1&2&1&5&1&4&1&9&1&2&1&3&1&2&1&7&1&4&1&6&1&2&1&3&1&2&1&5&1&4&1&9&1&2&1&3&1&2&1&12&1&4&1&7&1&2&1&3&1&2&1&5&1&4&1&8\\
15&1&5&1&7&1&3&1&2&1&3&1&9&1&5&1&4&1&3&1&2&1&3&1&6&1&5&1&13&1&3&1&2&1&3&1&8&1&5&1&4&1&3&1&2&1&3&1&7&1&5&1&6&1&3&1&2&1&3&1&9&1&5&1&4&1&3&1&2&1&3&1\\
1&3&1&2&1&4&1&11&1&5&1&2&1&3&1&2&1&12&1&10&1&7&1&2&1&3&1&2&1&4&1&11&1&5&1&2&1&3&1&2&1&6&1&11&1&14&1&2&1&3&1&2&1&4&1&10&1&5&1&2&1&3&1&2&1&7&1&11&1&6&1&2\\
4&1&9&1&3&1&2&1&3&1&7&1&4&1&6&1&3&1&2&1&3&1&5&1&4&1&8&1&3&1&2&1&3&1&12&1&4&1&7&1&3&1&2&1&3&1&5&1&4&1&8&1&3&1&2&1&3&1&6&1&4&1&13&1&3&1&2&1&3&1&5&1\\
1&2&1&6&1&5&1&14&1&2&1&3&1&2&1&8&1&5&1&4&1&2&1&3&1&2&1&7&1&5&1&6&1&2&1&3&1&2&1&9&1&5&1&4&1&2&1&3&1&2&1&15&1&5&1&7&1&2&1&3&1&2&1&8&1&5&1&4&1&2&1&3\\
7&1&3&1&2&1&3&1&4&1&10&1&5&1&3&1&2&1&3&1&6&1&11&1&14&1&3&1&2&1&3&1&4&1&10&1&5&1&3&1&2&1&3&1&7&1&10&1&6&1&3&1&2&1&3&1&4&1&11&1&5&1&3&1&2&1&3&1&12&1&10&1\\
1&5&1&4&1&8&1&2&1&3&1&2&1&15&1&4&1&7&1&2&1&3&1&2&1&5&1&4&1&9&1&2&1&3&1&2&1&6&1&4&1&13&1&2&1&3&1&2&1&5&1&4&1&9&1&2&1&3&1&2&1&7&1&4&1&6&1&2&1&3&1&2\\
3&1&2&1&3&1&7&1&5&1&6&1&3&1&2&1&3&1&9&1&5&1&4&1&3&1&2&1&3&1&15&1&5&1&7&1&3&1&2&1&3&1&8&1&5&1&4&1&3&1&2&1&3&1&6&1&5&1&14&1&3&1&2&1&3&1&9&1&5&1&4&1\\
1&11&1&13&1&2&1&3&1&2&1&4&1&11&1&5&1&2&1&3&1&2&1&7&1&10&1&6&1&2&1&3&1&2&1&4&1&11&1&5&1&2&1&3&1&2&1&12&1&11&1&7&1&2&1&3&1&2&1&4&1&10&1&5&1&2&1&3&1&2&1&6\\
2&1&3&1&5&1&4&1&9&1&3&1&2&1&3&1&6&1&4&1&8&1&3&1&2&1&3&1&5&1&4&1&8&1&3&1&2&1&3&1&7&1&4&1&6&1&3&1&2&1&3&1&5&1&4&1&8&1&3&1&2&1&3&1&15&1&4&1&7&1&3&1\\
1&4&1&2&1&3&1&2&1&12&1&5&1&7&1&2&1&3&1&2&1&13&1&5&1&4&1&2&1&3&1&2&1&6&1&5&1&14&1&2&1&3&1&2&1&9&1&5&1&4&1&2&1&3&1&2&1&7&1&5&1&6&1&2&1&3&1&2&1&8&1&5\\
3&1&7&1&10&1&6&1&3&1&2&1&3&1&4&1&10&1&5&1&3&1&2&1&3&1&12&1&11&1&7&1&3&1&2&1&3&1&4&1&10&1&5&1&3&1&2&1&3&1&6&1&10&1&13&1&3&1&2&1&3&1&4&1&11&1&5&1&3&1&2&1\\
1&2&1&3&1&2&1&5&1&4&1&8&1&2&1&3&1&2&1&7&1&4&1&6&1&2&1&3&1&2&1&5&1&4&1&9&1&2&1&3&1&2&1&15&1&4&1&7&1&2&1&3&1&2&1&5&1&4&1&12&1&2&1&3&1&2&1&6&1&4&1&9\\
14&1&5&1&4&1&3&1&2&1&3&1&6&1&5&1&14&1&3&1&2&1&3&1&9&1&5&1&4&1&3&1&2&1&3&1&7&1&5&1&6&1&3&1&2&1&3&1&8&1&5&1&4&1&3&1&2&1&3&1&9&1&5&1&7&1&3&1&2&1&3&1\\
1&3&1&2&1&15&1&11&1&7&1&2&1&3&1&2&1&4&1&11&1&5&1&2&1&3&1&2&1&6&1&10&1&13&1&2&1&3&1&2&1&4&1&11&1&5&1&2&1&3&1&2&1&7&1&11&1&6&1&2&1&3&1&2&1&4&1&10&1&5&1&2\\
4&1&6&1&3&1&2&1&3&1&5&1&4&1&9&1&3&1&2&1&3&1&15&1&4&1&7&1&3&1&2&1&3&1&5&1&4&1&8&1&3&1&2&1&3&1&6&1&4&1&9&1&3&1&2&1&3&1&5&1&4&1&8&1&3&1&2&1&3&1&7&1\\
1&2&1&8&1&5&1&4&1&2&1&3&1&2&1&7&1&5&1&6&1&2&1&3&1&2&1&8&1&5&1&4&1&2&1&3&1&2&1&12&1&5&1&7&1&2&1&3&1&2&1&14&1&5&1&4&1&2&1&3&1&2&1&6&1&5&1&13&1&2&1&3\\
5&1&3&1&2&1&3&1&6&1&10&1&13&1&3&1&2&1&3&1&4&1&10&1&5&1&3&1&2&1&3&1&7&1&11&1&6&1&3&1&2&1&3&1&4&1&10&1&5&1&3&1&2&1&3&1&15&1&10&1&7&1&3&1&2&1&3&1&4&1&11&1\\
1&12&1&4&1&7&1&2&1&3&1&2&1&5&1&4&1&12&1&2&1&3&1&2&1&6&1&4&1&9&1&2&1&3&1&2&1&5&1&4&1&9&1&2&1&3&1&2&1&7&1&4&1&6&1&2&1&3&1&2&1&5&1&4&1&9&1&2&1&3&1&2\\
3&1&2&1&3&1&9&1&5&1&4&1&3&1&2&1&3&1&8&1&5&1&7&1&3&1&2&1&3&1&14&1&5&1&4&1&3&1&2&1&3&1&6&1&5&1&13&1&3&1&2&1&3&1&8&1&5&1&4&1&3&1&2&1&3&1&7&1&5&1&6&1\\
1&10&1&5&1&2&1&3&1&2&1&7&1&11&1&6&1&2&1&3&1&2&1&4&1&11&1&5&1&2&1&3&1&2&1&15&1&10&1&7&1&2&1&3&1&2&1&4&1&11&1&5&1&2&1&3&1&2&1&6&1&11&1&14&1&2&1&3&1&2&1&4\\
2&1&3&1&6&1&4&1&8&1&3&1&2&1&3&1&5&1&4&1&9&1&3&1&2&1&3&1&7&1&4&1&6&1&3&1&2&1&3&1&5&1&4&1&8&1&3&1&2&1&3&1&12&1&4&1&7&1&3&1&2&1&3&1&5&1&4&1&15&1&3&1\\
1&7&1&2&1&3&1&2&1&14&1&5&1&4&1&2&1&3&1&2&1&6&1&5&1&13&1&2&1&3&1&2&1&8&1&5&1&4&1&2&1&3&1&2&1&7&1&5&1&6&1&2&1&3&1&2&1&9&1&5&1&4&1&2&1&3&1&2&1&8&1&5\\
3&1&4&1&11&1&5&1&3&1&2&1&3&1&15&1&10&1&7&1&3&1&2&1&3&1&4&1&10&1&5&1&3&1&2&1&3&1&6&1&11&1&14&1&3&1&2&1&3&1&4&1&10&1&5&1&3&1&2&1&3&1&7&1&10&1&6&1&3&1&2&1\\
1&2&1&3&1&2&1&7&1&4&1&6&1&2&1&3&1&2&1&5&1&4&1&8&1&2&1&3&1&2&1&12&1&4&1&7&1&2&1&3&1&2&1&5&1&4&1&15&1&2&1&3&1&2&1&6&1&4&1&8&1&2&1&3&1&2&1&5&1&4&1&9\\
6&1&5&1&13&1&3&1&2&1&3&1&9&1&5&1&4&1&3&1&2&1&3&1&7&1&5&1&6&1&3&1&2&1&3&1&9&1&5&1&4&1&3&1&2&1&3&1&9&1&5&1&7&1&3&1&2&1&3&1&13&1&5&1&4&1&3&1&2&1&3&1\\
1&3&1&2&1&4&1&10&1&5&1&2&1&3&1&2&1&6&1&11&1&14&1&2&1&3&1&2&1&4&1&11&1&5&1&2&1&3&1&2&1&7&1&10&1&6&1&2&1&3&1&2&1&4&1&11&1&5&1&2&1&3&1&2&1&12&1&11&1&7&1&2\\
4&1&8&1&3&1&2&1&3&1&12&1&4&1&7&1&3&1&2&1&3&1&5&1&4&1&15&1&3&1&2&1&3&1&6&1&4&1&8&1&3&1&2&1&3&1&5&1&4&1&8&1&3&1&2&1&3&1&7&1&4&1&6&1&3&1&2&1&3&1&5&1\\
1&2&1&7&1&5&1&6&1&2&1&3&1&2&1&8&1&5&1&4&1&2&1&3&1&2&1&9&1&5&1&7&1&2&1&3&1&2&1&13&1&5&1&4&1&2&1&3&1&2&1&6&1&5&1&14&1&2&1&3&1&2&1&9&1&5&1&4&1&2&1&3\\
14&1&3&1&2&1&3&1&4&1&11&1&5&1&3&1&2&1&3&1&7&1&10&1&6&1&3&1&2&1&3&1&4&1&10&1&5&1&3&1&2&1&3&1&12&1&11&1&7&1&3&1&2&1&3&1&4&1&10&1&5&1&3&1&2&1&3&1&6&1&10&1\\
1&5&1&4&1&15&1&2&1&3&1&2&1&6&1&4&1&13&1&2&1&3&1&2&1&5&1&4&1&8&1&2&1&3&1&2&1&7&1&4&1&6&1&2&1&3&1&2&1&5&1&4&1&9&1&2&1&3&1&2&1&15&1&4&1&7&1&2&1&3&1&2\\
3&1&2&1&3&1&9&1&5&1&7&1&3&1&2&1&3&1&9&1&5&1&4&1&3&1&2&1&3&1&6&1&5&1&14&1&3&1&2&1&3&1&9&1&5&1&4&1&3&1&2&1&3&1&7&1&5&1&6&1&3&1&2&1&3&1&8&1&5&1&4&1\\
1&11&1&6&1&2&1&3&1&2&1&4&1&10&1&5&1&2&1&3&1&2&1&12&1&11&1&7&1&2&1&3&1&2&1&4&1&11&1&5&1&2&1&3&1&2&1&6&1&10&1&13&1&2&1&3&1&2&1&4&1&11&1&5&1&2&1&3&1&2&1&7\\
2&1&3&1&5&1&4&1&8&1&3&1&2&1&3&1&7&1&4&1&6&1&3&1&2&1&3&1&5&1&4&1&9&1&3&1&2&1&3&1&15&1&4&1&7&1&3&1&2&1&3&1&5&1&4&1&12&1&3&1&2&1&3&1&6&1&4&1&9&1&3&1\\
1&4&1&2&1&3&1&2&1&6&1&5&1&14&1&2&1&3&1&2&1&8&1&5&1&4&1&2&1&3&1&2&1&7&1&5&1&6&1&2&1&3&1&2&1&8&1&5&1&4&1&2&1&3&1&2&1&8&1&5&1&7&1&2&1&3&1&2&1&13&1&5\\
3&1&12&1&10&1&7&1&3&1&2&1&3&1&4&1&11&1&5&1&3&1&2&1&3&1&6&1&10&1&13&1&3&1&2&1&3&1&4&1&10&1&5&1&3&1&2&1&3&1&7&1&11&1&6&1&3&1&2&1&3&1&4&1&10&1&5&1&3&1&2&1\\
1&2&1&3&1&2&1&5&1&4&1&9&1&2&1&3&1&2&1&15&1&4&1&7&1&2&1&3&1&2&1&5&1&4&1&12&1&2&1&3&1&2&1&6&1&4&1&9&1&2&1&3&1&2&1&5&1&4&1&9&1&2&1&3&1&2&1&7&1&4&1&6\\
8&1&5&1&4&1&3&1&2&1&3&1&7&1&5&1&6&1&3&1&2&1&3&1&9&1&5&1&4&1&3&1&2&1&3&1&8&1&5&1&7&1&3&1&2&1&3&1&14&1&5&1&4&1&3&1&2&1&3&1&6&1&5&1&14&1&3&1&2&1&3&1\\
1&3&1&2&1&6&1&11&1&13&1&2&1&3&1&2&1&4&1&10&1&5&1&2&1&3&1&2&1&7&1&11&1&6&1&2&1&3&1&2&1&4&1&11&1&5&1&2&1&3&1&2&1&15&1&10&1&7&1&2&1&3&1&2&1&4&1&11&1&5&1&2\\
4&1&7&1&3&1&2&1&3&1&5&1&4&1&8&1&3&1&2&1&3&1&6&1&4&1&8&1&3&1&2&1&3&1&5&1&4&1&9&1&3&1&2&1&3&1&7&1&4&1&6&1&3&1&2&1&3&1&5&1&4&1&8&1&3&1&2&1&3&1&15&1\\
1&2&1&9&1&5&1&4&1&2&1&3&1&2&1&12&1&5&1&7&1&2&1&3&1&2&1&14&1&5&1&4&1&2&1&3&1&2&1&6&1&5&1&13&1&2&1&3&1&2&1&8&1&5&1&4&1&2&1&3&1&2&1&7&1&5&1&6&1&2&1&3\\
5&1&3&1&2&1&3&1&7&1&10&1&6&1&3&1&2&1&3&1&4&1&11&1&5&1&3&1&2&1&3&1&15&1&10&1&7&1&3&1&2&1&3&1&4&1&10&1&5&1&3&1&2&1&3&1&6&1&11&1&13&1&3&1&2&1&3&1&4&1&10&1\\
1&6&1&4&1&14&1&2&1&3&1&2&1&5&1&4&1&9&1&2&1&3&1&2&1&7&1&4&1&6&1&2&1&3&1&2&1&5&1&4&1&8&1&2&1&3&1&2&1&12&1&4&1&7&1&2&1&3&1&2&1&5&1&4&1&9&1&2&1&3&1&2\\
3&1&2&1&3&1&8&1&5&1&4&1&3&1&2&1&3&1&6&1&5&1&13&1&3&1&2&1&3&1&9&1&5&1&4&1&3&1&2&1&3&1&7&1&5&1&6&1&3&1&2&1&3&1&9&1&5&1&4&1&3&1&2&1&3&1&12&1&5&1&7&1\\
1&11&1&5&1&2&1&3&1&2&1&15&1&11&1&7&1&2&1&3&1&2&1&4&1&10&1&5&1&2&1&3&1&2&1&6&1&11&1&14&1&2&1&3&1&2&1&4&1&11&1&5&1&2&1&3&1&2&1&7&1&10&1&6&1&2&1&3&1&2&1&4\\
2&1&3&1&7&1&4&1&6&1&3&1&2&1&3&1&5&1&4&1&8&1&3&1&2&1&3&1&12&1&4&1&7&1&3&1&2&1&3&1&5&1&4&1&9&1&3&1&2&1&3&1&6&1&4&1&8&1&3&1&2&1&3&1&5&1&4&1&8&1&3&1\\
1&13&1&2&1&3&1&2&1&9&1&5&1&4&1&2&1&3&1&2&1&7&1&5&1&6&1&2&1&3&1&2&1&8&1&5&1&4&1&2&1&3&1&2&1&15&1&5&1&7&1&2&1&3&1&2&1&14&1&5&1&4&1&2&1&3&1&2&1&6&1&5\\
3&1&4&1&10&1&5&1&3&1&2&1&3&1&6&1&10&1&14&1&3&1&2&1&3&1&4&1&11&1&5&1&3&1&2&1&3&1&7&1&10&1&6&1&3&1&2&1&3&1&4&1&10&1&5&1&3&1&2&1&3&1&15&1&11&1&7&1&3&1&2&1\\
\end{tabular}
}
\end{center}
\caption{Our $72 \times 72$ periodic solution in $15$ colours}
\label{fig:15}
\end{figure}
\end{landscape}

\section{Discussion}

We believe Simulated Annealing might well not have found our $16$-colouring solution and one argument is furnished for this by the tables in Figure~\ref{fig:frequency}. For the Soukal-Holub $17$-colouring, the frequency of each colour monotonely decreases as the colour number rises (something quite unsurprising). Whereas for our $16$-colouring, this monotonicity is broken in two places. Simulated Annealing allows for some suboptimal searching, but the algorithm from \cite{SoukalHolub10} is greedy colour-by-colour, in ascending order, and so it seems rather unlikely that it would throw this break of monotonicity. This may be advanced as an argument in favour of SAT-solving for this sort of problem, so long as the search space is suitably restricted. However, we note our $15$-colouring is monotone. 

In Figure~\ref{fig:running-times}, we can see running times for certain instances with the basic encoding versus encoding using commander variables. The designation, for example, 72x72-15(1-5) indicates the grid was $72 \times 72$ planted with colours $1$ to $5$ and we were looking for a $15$-colouring. One can see that the running times were significantly improved by the addition of commander variables. This improvement is unsurprising given what is known about the Pigeonhole Principle, which performs better under the commander encoding, and which has a flavour not dissimilar to our problem (see \cite{Commander}, also \cite{HN2013} and Chapter 7 in \cite{JustynaBook}). Note that commander encodings are a well-known heuristic in the SAT-solving community, but it is not surely known why they lead to more rapid solutions. Indeed, there is no reason a priori for thinking they will always lead to faster solutions.

We note that the $15$-colouring was somewhat easier to find on the $72 \times 72$ grid than the $16$-colouring on the $48 \times 48$ grid. We do not have an intuition as to why this might be.

\begin{figure}
\begin{center}
\begin{tabular}{|c|c||c|c|}
\hline
& SAT? & basic G Syrup & cmdr G Syrup \\
\hline
72x72-14(1-4)$^*$ & unsat & 4913 & 1290 \\
\hline
72x72-15(1-5) & sat & 464 & 420 \\
\hline
48x48-16(1-8) & unsat & 112 & 35\\
\hline
48x48-16(1-7)$^*$  & sat & 18589 & 10519 \\
\hline
24x24-16(1-7)  & unsat & 82 & 35 \\
\hline
\end{tabular}
\end{center}
\caption{Running times in seconds on interesting instances}
\label{fig:running-times}
\end{figure}

\section*{Acknowledgements}

The authors are grateful to Bernard Lidick\'y and Roman Soukal for many discussions on the problem. We are also grateful to several anonymous reviewers for corrections and clarifications.

\bibliographystyle{acm}

\end{document}